\documentclass[journal]{IEEEtran}
\usepackage{cite}
\usepackage{amsmath,amssymb,amsfonts}
\usepackage{graphicx}
\usepackage{algorithm}
\usepackage{algpseudocode}
\usepackage{array}
\usepackage{balance}
\usepackage[hidelinks]{hyperref}

\begin{document}

\title{Conditional-Coverage Contributor Selection for Regional Digital
Twins Under Mobility}

\author{Tao~Yu%
\thanks{T. Yu is with the Institute of Science Tokyo, Tokyo, Japan
(e-mail: yutao@mobile.ee.titech.ac.jp).}}

\markboth{IEEE Networking Letters}%
{Yu: Conditional-Coverage Contributor Selection for Regional Digital Twins Under Mobility}

\maketitle

\begin{abstract}
Regional digital twins (DTs) under mobility must coordinate contributor admission over congested broadcast networks without fixed infrastructure. Per-sender redundancy mitigation cannot make this set-level decision because only the host has a region-wide coverage map. Treating regional state as a public good, this letter develops a host-coordinated protocol admitting contributors whose timely, non-redundant coverage gain exceeds a congestion price within a cycle deadline. A 13-byte beacon digest enables pre-transmission scoring. Standards-based simulation shows the protocol reduces the timely actionable coverage deficit from 5.6\% to 1.4\% at load comparable to ETSI redundancy mitigation and by 26\% versus random admission at matched tight-budget load.
\end{abstract}


\section{Introduction}\label{sec:intro}
Digital twins (DTs) have emerged as a practical paradigm for connecting teams of mobile agents, such as vehicles, drones, and robot fleets, to a live representation of the region they operate in. The fused regional state serves whatever tasks the agents run and lets each participant act on conditions beyond its own observation~\cite{Yu2024IoFDT,Wang2024SMDTNav}. Most existing designs, however, assume fixed geographic scopes, fixed information sources, and infrastructure-hosted fusion points. Under mobility these assumptions fail, since the set of agents that can observe a region, the set that needs the fused state, and the agent best placed to host the fusion all change within seconds. Sustaining such a twin is a network-control problem. Periodic beacons alone claim much of the shared channel, and nearby observations overlap heavily, so which agents should actively contribute is a central design question for any regional DT.

In vehicular networks, the ETSI collective perception service lets vehicles share their detected objects in collective perception messages (CPMs), with redundancy mitigation performed independently at senders~\cite{TR103562,TS103324,Lyu2025RedundancySurvey}, so the decision is per sender and per object. Helper selection~\cite{Sarlak2025} chooses which neighbors share raw sensor data to improve the perception of a single ego vehicle. Value-of-information control~\cite{Higuchi2019VoI} decides what a fixed sender includes in its report. The decision made here differs on three axes. It is made for a set, not per sender. It serves every receiver in a region, not one ego vehicle. And it decides which senders transmit at all, not what a given sender includes. Learned collaborative perception~\cite{Hu2022Where2comm} selects partners or spatial regions under bandwidth limits, but ego-centrically and with training, whereas the present decision is training-free and prices airtime on a standardized channel. DT frameworks for mobility organize fusion at cloud or edge infrastructure with a fixed scope and membership~\cite{Yu2024IoFDT,Wang2024SMDTNav}, whereas here the selected set, the delivered state, and the beneficiaries all move on a shared broadcast channel.

These approaches leave the collective decision unanswered. Per-sender mitigation suffers a structural information asymmetry: what matters is what each receiver lacks, and only a fusion point holding the group's coverage map knows it. Moreover, the fused DT state is a regional information public good. Every agent in the footprint receives the broadcast whether or not it contributed, so the benefit diffuses over the service scope while the cost falls on the contributor set. An agent whose view is already covered should therefore not be admitted, however close or task-critical it is.

Contributor admission is therefore modeled as a marginal decision of the host. An agent is admitted only when the timely, non-redundant coverage it adds, conditioned on everything the twin already receives, exceeds the regional airtime price of its uplink, within a hard service-cycle deadline. A compact coverage digest in the periodic beacon makes this computable before anything expensive is transmitted. This casts active DT membership as a host-coordinated contributor-admission protocol for a moving service scope, combining conditional coverage value, a regional congestion price, and hysteresis. It is instantiated with the ETSI cooperative awareness message (CAM) as beacon and the CPM as report, and validated in a standards-based corridor simulation.

\section{System Model}\label{sec:model}

\subsection{Scopes and Roles}
\begin{figure}[t]
\centering
\includegraphics[width=0.9\columnwidth]{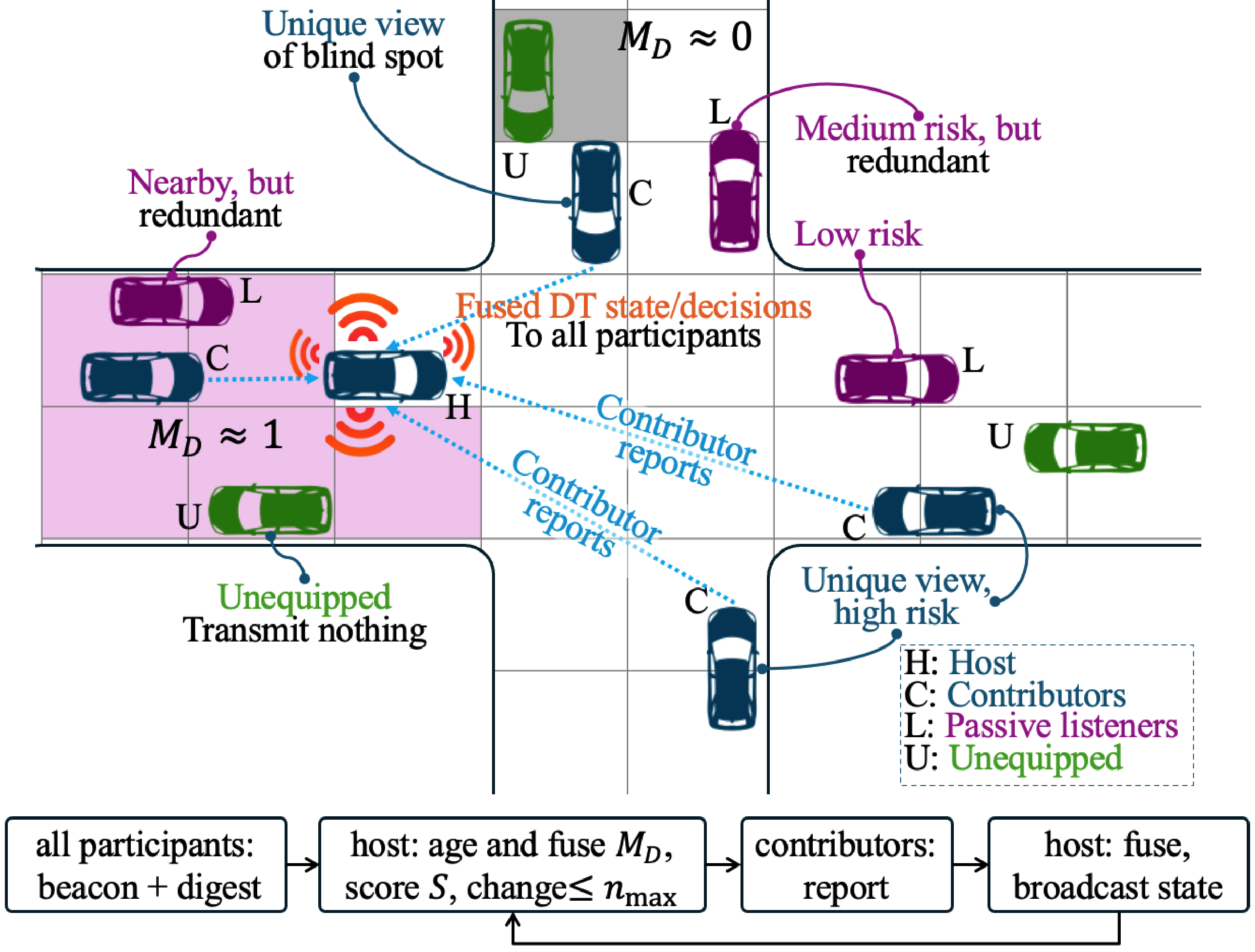}
\caption{Host-coordinated contributor selection in a mobile regional DT.
Top: host H fuses reports from admitted contributors C and broadcasts fused states to all participants, including passive listeners L. The overlay sketches the coverage grid $\mathcal{G}$ (not to scale). Cells the twin already holds have $M_{\mathcal{D}}\approx1$, and the occluded cell hiding unequipped vehicle U has $M_{\mathcal{D}}\approx0$ until the top contributor is admitted for its unique view of it. Redundant candidates are not admitted however nearby or risk-exposed. Bottom: one service cycle $T_s$.}
\label{fig:concept}
\end{figure}

Consider a mobile broadcast network in a bounded region. The following terms are used throughout (Fig.~\ref{fig:concept}). An \emph{agent} is any mobile node. A \emph{participating agent} carries sensors and a radio, broadcasts a periodic beacon (the CAM here), and runs a local task whose quality depends on timely knowledge of the regional state. \emph{Objects} are everything the twin represents. An unequipped vehicle is a non-participating agent that transmits nothing and is represented only as an object. The map is partitioned into fixed square tiles, the \emph{resource regions} $\mathcal{R}$. Each tile carries one regional DT $k$ with one \emph{host} $h$, a participating agent chosen by the beacon-based rule of Sect.~\ref{sec:algorithm}. The host fuses reports and broadcasts the fused state. Several regional DTs thus coexist over adjacent tiles, and a participant may overhear more than one host. Each regional DT $k$ maintains two sets. The \emph{contributor scope} $\mathcal{D}_k(t)$ comprises the host and the \emph{contributors}, the participating agents currently admitted to upload detailed reports (CPMs) at the cost of uplink airtime and fusion resources. The \emph{service scope} $\mathcal{F}_k(t)$ contains all participating agents inside the host's current broadcast footprint, which receive fused DT state whether or not they contribute. Its non-contributing members are the \emph{passive listeners}, which form the \emph{candidate set} $\mathcal{J}=\mathcal{F}_k\setminus\mathcal{D}_k$ for admission.

The host fuses contributor reports in each cycle of fixed period $T_s$ and broadcasts the fused state to $\mathcal{F}_k$. The airtime that the uplinks and the broadcast claim within one cycle, $T_{\mathrm{cyc}}(\mathcal{D}_k,h)$, grows with the contributor set. The twin's benefit is thus summed over $\mathcal{F}_k$ while its cost is borne by $\mathcal{D}_k$ alone. This asymmetry separates two properties that are easily conflated. \emph{Task relevance} is how much an agent needs the fused state, set by its exposure to conflict, and it fixes the agent's weight in the benefit sum. \emph{Contributor eligibility} is whether the agent should be admitted to $\mathcal{D}_k$, and it depends only on what its observations add to what the twin already holds. A vehicle entering a blind intersection may be highly relevant yet ineligible because the cells it sees are already covered, while a low-risk vehicle may be the only one that sees an occluded lane (Fig.~\ref{fig:concept}). Because the host is itself a participating agent, control is centralized within each region and distributed across regions, with no roadside unit or cloud. Agents are assumed cooperative, and tasks enter only through a spatial risk prior and per-object reaction deadlines, both instantiated in Sect.~\ref{sec:instantiation}.

\subsection{Information Available for Admission}
{With contributor set $\mathcal{D}$, the information available to agent $i$ is
\begin{equation}
\mathcal{I}_i(\mathcal{D},t) = \mathcal{I}_i^{\mathrm{sens}}(t)
\cup \mathcal{I}_i^{\mathrm{bcn}}(t)
\cup \mathcal{I}_i^{\mathrm{oh}}(t) \cup
\textstyle\bigcup_{m\in\mathcal{D}} \mathcal{P}_m(t)
\label{eq:fullinfo}
\end{equation}
which comprises local sensing under occlusion, received beacons, overheard DT-state broadcasts from one or more hosts, and the detailed reports $\mathcal{P}_m(t)$ of the contributors as delivered to $i$ by the transport.} The \emph{coverage grid} $\mathcal{G}$ partitions the map into square cells of side 20\,m aligned to a global origin, so that every host indexes cells identically. A cell is \emph{visible} to an agent when it lies within sensing range and its line of sight is not blocked by buildings or vehicles. In addition to the sender state, each beacon carries a \emph{coverage digest}. Its 12-byte header holds the container identifier and length (legacy receivers skip it under ASN.1 extensibility), the absolute grid anchor, and an encoding flag. It is followed by a variable-length run-length encoding of the bitmap of the sender's currently visible cells. Under heavy occlusion by buildings and surrounding vehicles the bitmap is sparse, and its measured average is about 1\,B, giving a digest of about 13\,B. It reveals where an agent can observe without revealing object states, so a candidate's value is computable before admission and no detailed information is given away beforehand.

\subsection{Conditional Incremental Coverage Value}
The value of admitting candidate $j$ is the task-loss reduction it induces across the service scope, conditioned on all information already available,
\begin{align}
\Delta V(j\mid\mathcal{D},t) = \sum_{i\in\mathcal{F}}
\big( &\mathbb{E}[L_i \mid \mathcal{I}_i(\mathcal{D},t)] \nonumber\\
&- \mathbb{E}[L_i \mid \mathcal{I}_i(\mathcal{D},t)\cup\mathcal{P}_j(t)]\big)
\label{eq:deltaV}
\end{align}
where $L_i$ is the task loss of agent $i$. Here $L_i$ is instantiated as the mass of decision-relevant conflict risk that agent $i$ fails to know in time, i.e., the risk-weighted count of objects on a conflict course with $i$ whose state $i$ has not received within its reaction deadline. The task-loss reduction in \eqref{eq:deltaV} is the amount by which the report of $j$ shrinks this unknown-in-time mass, summed over the service scope. The summation runs over $\mathcal{F}$ rather than $\mathcal{D}$ because the beneficiary of a filled blind spot is often a passive listener. Eq.~\eqref{eq:deltaV} is estimated on $\mathcal{G}$ as
\begin{equation}
\widehat{\Delta V}_j = \sum_{g\in\mathcal{G}}
\big(1 - M_{\mathcal{D}}(g)\big)\,
q_j(g)\,
\rho_{\mathcal{F}}(g)\,
P\big(A_g \le \Delta^{\mathrm{react}}_g\big)
\label{eq:score}
\end{equation}
The four factors represent novelty, observability, task risk, and timely delivery. Motivated by \eqref{eq:deltaV}, this factorized surrogate assumes additive cell losses and conditional independence. $M_{\mathcal{D}}(g)\in[0,1]$ is a soft coverage-confidence map that the host fuses from admitted contributors' digests and reports plus overheard state, $q_j(g)\in\{0,1\}$ is the visibility of cell $g$ read from the digest of candidate $j$, and $\rho_{\mathcal{F}}$ aggregates map-derived conflict risk over the predicted paths of service-scope members. The last factor is the probability that the information age $A_g$ at the receiver stays within the reaction deadline $\Delta^{\mathrm{react}}_g$ given the post-admission completion time $T_{\mathrm{cyc}}$, i.e., the complement of an age-of-information violation probability~\cite{Yates2021AoI}. The score credits only delivery through the host broadcast, and it decreases as the post-admission $T_{\mathrm{cyc}}$ grows, coupling value to the set size. Novelty covers both completion of known objects and never-observed areas under one deduplication rule, and all factors are evaluated at the digest resolution, a finer grid serving only ground-truth evaluation.

{The host maintains $M_{\mathcal{D}}$ recursively. Each cycle it ages the previous map, $\bar M(g,t)=e^{-\kappa T_s}M_{\mathcal{D}}(g,t{-}T_s)$, and fuses the coverage evidence of the current cycle as independent events,
\begin{equation}
M_{\mathcal{D}}(g,t) = 1 - \big(1 - \bar M(g,t)\big)
\prod_{m\in\mathcal{D}}\big(1 - c_m(g)\big)
\prod_{o\in\mathcal{O}}\big(1 - I_o(g)\big)
\label{eq:mdupdate}
\end{equation}
where $c_m(g)=q_m(g)$ if the report of contributor $m$ was decoded in cycle $t$ and $c_m(g)=p_m\,q_m(g)$ otherwise, with $p_m$ the decoded fraction of $m$'s reports over a sliding window of recent cycles and $p_h=1$ for the host's own sensing, and $I_o(g)=1$ if a broadcast overheard in cycle $t$ from a neighboring host $o$, with $\mathcal{O}$ the set of such hosts, places an object with a current-cycle timestamp in $g$. A decoded report thus removes the delivery discount $p_m$ on the digest's expected coverage, so no source is counted twice, and the time constant $1/\kappa=0.5$\,s lies within the range of reaction deadlines (Table~\ref{tab:params}).
}

\subsection{Network Load and Regional Congestion Control}
With orthogonal uplinks the airtime of a regional DT is
\begin{equation}
C_{\mathrm{air}}(\mathcal{D},h) = \sum_{m\in\mathcal{D}\setminus\{h\}}
\frac{B_m}{R_{\mathrm{phy}}} + C_{\mathrm{bc}}(\mathcal{D},h)
\label{eq:cair}
\end{equation}
where $B_m$ is the report size, $R_{\mathrm{phy}}$ the configured physical rate, and $C_{\mathrm{bc}}$ the airtime of the fused-state broadcast. This orthogonal accounting abstracts the distributed scheduling of the radio. Contention and collisions enter through the channel model of Sect.~\ref{sec:eval}. Two load controls act on each host. The \emph{per-host allowance} $b$ caps the fraction of channel capacity that one host's twin may consume per cycle through \eqref{eq:cair}. It is a hard bound the host enforces alone, treated as a design parameter and swept in Sect.~\ref{sec:eval}. The \emph{regional congestion target} $\Gamma_{\mathcal{R}}$ caps the total occupancy of the tile from all sources, including traffic the host does not control. It is enforced softly through a regional congestion price updated by projected dual ascent, $\lambda_{\mathcal{R}}\!\leftarrow\![\lambda_{\mathcal{R}}+\mu(\hat c_{\mathcal{R}}-\Gamma_{\mathcal{R}})]_+$ with $\mu{=}0.5$ each beacon interval and $\hat c_{\mathcal{R}}$ the locally measured occupancy, which prices every admission. Mandatory beacons are excluded from this load because admission cannot change them. A hard constraint $T_{\mathrm{cyc}}(\mathcal{D},h)\le T_{\max}$ guarantees real-time feasibility. The two controls bind in different regimes: with a small $b$ the allowance is reached first and $\lambda_{\mathcal{R}}$ stays near zero, whereas with a generous $b$ the price rises until the offered load settles just below $\Gamma_{\mathcal{R}}$ (Sect.~\ref{sec:results}). Classical congestion pricing regulates source rates through shadow prices~\cite{Kelly1998}, whereas here the price is weighed against the conditional information gain of a moving source.

\section{Host-Coordinated Contributor Admission}\label{sec:algorithm}

In each service cycle the incumbent host is retained while it remains in the resource region. Otherwise the participating agent nearest the region center is selected, using beacon-reported positions with a deterministic tie-break, which anchors the broadcast footprint to the fixed tile. From digests, decoded reports, overheard state, and link statistics, the host updates $M_{\mathcal{D}}$ by \eqref{eq:mdupdate}, the service scope, and the risk map. Each candidate $j\in\mathcal{J}$ is scored by the net marginal value
\begin{equation}
S(j\mid\mathcal{D}) = \widehat{\Delta V}_j -
\lambda_{\mathcal{R}}\,\Delta C_{\mathrm{air}}(j\mid\mathcal{D})
\label{eq:netscore}
\end{equation}
A candidate is admitted when $S(j\mid\mathcal{D})\ge\eta_{\mathrm{add}}$ and the post-admission cycle satisfies the cycle constraint and the per-host allowance $b$. An existing contributor is removed when its retention score drops to $\eta_{\mathrm{rem}}$ or below, where $\eta_{\mathrm{rem}} < \eta_{\mathrm{add}}$ creates hysteresis against oscillation. The symmetric hysteresis half-width $x$ ($\eta_{\mathrm{add}}{=}x$, $\eta_{\mathrm{rem}}{=}{-}x$) exists to suppress oscillation, not to tune performance, and was fixed on ten calibration seeds disjoint from the evaluation seeds, as a tenth of the median positive retention score of admitted members ($x=0.25$), before any performance metric was inspected. In operation the thresholds mainly filter out near-zero-value candidates, and selection is performed by the score ordering. Admissions and removals share a per-cycle change quota $n_{\max}$ ($=1$ in the evaluation, with scores recomputed after any committed change), so the set evolves by single marginal steps separated by the hysteresis gap. The thresholds decide only when the best candidate lies within the dead band. At small $b$ the allowance is saturated and at large $b$ the price term dominates the score, so $x$ was not swept. Thresholds adapting to local density or speed are future work. Each committed change advances a logical epoch.

With $M_{\mathcal{D}}$ as a group-level surrogate, the host evaluates the grid in $O((|\mathcal{J}|{+}|\mathcal{F}|)|\mathcal{G}|)$ and sorts candidates in $O(|\mathcal{J}|\log|\mathcal{J}|)$ per cycle.

\begin{algorithm}[t]
\caption{Contributor admission in each cycle $T_s$}
\label{alg:adapt}
\begin{algorithmic}[1]
\State update $M_{\mathcal{D}}$ by \eqref{eq:mdupdate}, $\mathcal{F}$, $\rho_{\mathcal{F}}$, link statistics, and $\lambda_{\mathcal{R}}$
\State score candidates $\mathcal{J}$ and current members by \eqref{eq:netscore}
\State for $j\in\mathcal{J}$ in descending score, at most $n_{\max}$ changes: admit
  ($\mathcal{D}\gets\mathcal{D}\cup\{j\}$) if $S(j\mid\mathcal{D})\ge\eta_{\mathrm{add}}$,
  $T_{\mathrm{cyc}}(\mathcal{D}\cup\{j\})\le T_{\max}$, and allowance $b$ is feasible
\State remove non-host members with $S(m\mid\mathcal{D}\setminus\{m\})\le\eta_{\mathrm{rem}}$,
in ascending score, within the same $n_{\max}$ quota, and advance the epoch
\end{algorithmic}
\end{algorithm}

\section{Evaluation in Cooperative Mobility}\label{sec:eval}

\subsection{Vehicular Instantiation}\label{sec:instantiation}
The framework maps onto the vehicular standards landscape with a single new element. The beacon is the ETSI CAM (EN~302~637-2) at 10\,Hz, extended by the coverage digest as its only non-standard container. The CAM is the only message every participant emits, so the digest cannot ride the CPM, which candidates do not send. The detailed report is a CPM (TS~103~324), generated once per service cycle of $T_s=0.2$\,s by admitted contributors only. The shared channel is LTE-V2X (C-V2X) PC5 mode~4, and the regional congestion target is expressed in its subchannel airtime. The risk prior $\rho$ follows from map conflict geometry and the predicted paths of service-scope members, and the reaction deadline follows from emergency-braking kinematics.

The protocol runs above standard CAM/CPM dissemination and PC5 delivery. A contributor report is an ordinary broadcast CPM, received and fused directly by every vehicle in range exactly as under the per-sender baseline. The host broadcast supplements this one-hop channel, reaching beyond the sender's range and carrying changed object states, including unequipped vehicles, each cycle plus a full refresh every 1\,s, so the DT state is soft state with bounded staleness. The selection decides which few vehicles transmit at all. Listeners deduplicate multiple hosts by timestamp, confidence, and a deterministic tie-break. Because the DT state is soft, host replacement needs no handover: a host that leaves the tile is replaced by the rule of Sect.~\ref{sec:algorithm}, and the new host builds $M_{\mathcal{D}}$ from the digests it already receives and from overheard state, whose 1-s refresh bounds the initial staleness. The nearest-center rule anchors the footprint to the tile rather than to a vehicle, so under high mobility the host changes without moving the service scope, whereas under strongly non-uniform traffic it may pick a host at the platoon edge. Stability-aware cluster-head rules~\cite{Cooper2017Clustering} or learning-based selection~\cite{Wu2026AgenticSurvey} could refine host placement, which, with migration and cross-host consistency, is out of scope.

\subsection{Setup}
The simulation uses an OpenStreetMap-extracted signalized corridor (Tamagawa-d\={o}ri, Shibuya, Tokyo, 3.7\,km, 22 signalized intersections). SUMO traffic is held by closed-population control at the congested density $1.5\,k_c$, where $k_c=28.72$\,veh/km/lane is measured from the corridor's fundamental
diagram, with 80\,\% connected penetration. Per-host allowances
$b\in\{0.2, 0.4, 0.8\}$ of capacity are swept under a fixed regional congestion target $\Gamma_{\mathcal{R}}=0.3$. The cooperative load counts all airtime beyond the mandatory base CAM, namely digest extensions, contributor reports, and host broadcasts, against the full channel capacity. Building footprints from the same extract drive both sensor occlusion and radio NLOS states.

Maximum-rate CAMs (10\,Hz, a worst case with congestion control disabled)
alone occupy 64\% of capacity on average, digests excluded, so report
airtime is scarce by measurement. The digest adds $\approx$0.07 capacity units, charged to the proposed scheme's cooperative load. The channel abstracts PC5 mode~4 with the TR~37.885 urban link states, published block-error curves, load-dependent collision loss, and hard capacity enforcement. All message types count toward regional occupancy, though this load-level abstraction does not separate sender-count effects (Table~\ref{tab:params}).

\begin{table}[t]
\caption{Key simulation parameters.}
\label{tab:params}
\centering
\scriptsize
\setlength{\tabcolsep}{2pt}
\renewcommand{\arraystretch}{1.0}
\begin{tabular}{
>{\raggedright\arraybackslash}p{0.29\columnwidth}
>{\raggedright\arraybackslash}p{0.53\columnwidth}
>{\raggedright\arraybackslash}p{0.13\columnwidth}}
\hline
\textbf{Parameter} & \textbf{Value} & \textbf{Source}\\
\hline
Step, CAM, cycle $T_s$ & 0.05, 0.1, 0.2\,s & \cite{EN302637}\\
CAM size & 121\,B (no security envelope) & \cite{EN302637}\\
Sensing range, FoV & 150\,m, $360^\circ$, ideal in LoS & \cite{TR103562}\\
Occlusion & 3D vehicle boxes, OSM buildings & \cite{TR103562,TR37885}\\
Pathloss, link states & urban LOS, NLOSv, NLOS & \cite{TR37885}\\
PHY & 5.9\,GHz, 10\,MHz, 2 subch., QPSK $r{=}0.5$ & \cite{TR37885,Molina2017}\\
BLER, load-dep.\ loss & published curves, $\alpha(\mathrm{CBR})$ & \cite{Molina2017}\\
Capacity enforcement & thinning $\min(1, C/L)$, hard cap & this work\\
Grids (eval., digest), tile & 5\,m, 20\,m ($\approx$13\,B @ CAM rate), 500\,m & design\\
Decay $\kappa$, link stat.\ $p_m$ & $2\,\mathrm{s^{-1}}$ ($1/\kappa{=}0.5$\,s), decoded fraction & design\\
Reaction deadline & $\tau_r{+}v_{\mathrm{rel}}/(2a_{\mathrm{br}})$, $\tau_r{=}0.2$\,s, $a_{\mathrm{br}}{=}9$\,m/s$^2$ & derived\\
$\eta_{\mathrm{add}}$, $\eta_{\mathrm{rem}}$, $n_{\max}$ & $+0.25$, $-0.25$ (10 calib.\ seeds), 1 & calib.\\
Density, penetration & $1.5\,k_c$ (measured), 80\,\% & scenario\\
Allowance $b$, target $\Gamma_{\mathcal{R}}$ & \{0.2, 0.4, 0.8\}, 0.3 of capacity & factor\\
\hline
\end{tabular}
\vspace{-1.0em}
\end{table}

The primary metric, the \emph{timely actionable coverage deficit}, is the pooled ratio, over all simulated vehicles and cycles, of decision-relevant objects unknown to a vehicle or staler than its reaction deadline to all decision-relevant objects. An object is decision-relevant to a vehicle when its ground-truth state lies on a map-derived conflict course with that vehicle's predicted path. Relevance is computed identically for all strategies and only for evaluation. The resource metric is the cooperative offered load normalized by channel capacity, booked when a message is offered. All strategies share paired seeds with identical traffic, equipment
assignment, and loss realizations. Results are reported with 95\% seed-level bootstrap
confidence intervals over 17 paired seeds, with 40\,s warm-up and 40\,s measurement.

\subsection{Baselines}
All baselines share the entire stack (sensing, CAM, channel, fusion, passive listening) and differ only in who reports and what is filtered. \textit{CPM-ETSI} follows the message generation rules of the collective perception service~\cite{TS103324} without optional redundancy mitigation, an unbudgeted load envelope. \textit{CPM-RMR} additionally applies the redundancy mitigation of TR~103~562: object self-announcement, frequency-based omission ($N_{\mathrm{RED}}{=}1$), and dynamics-based omission (4\,m, 0.5\,m/s), always active here (CBR gate 0.4), plus a cap of 32 objects per CPM ranked by novelty. \textit{Random top-$k$} inherits the hard cycle and per-host allowance constraints of Algorithm~\ref{alg:adapt} but replaces score-based admission with random choice, so it does not respond to $\lambda_{\mathcal{R}}$. Three criterion ablations keep the full stack and allowance but truncate the score \eqref{eq:score} to unconditional coverage ranking, to conditioning alone, and to conditioning with risk weighting. An oracle-awareness variant with lossless and delay-free delivery of all equipped observations provides the information ceiling, and CAM-only is the no-cooperation reference.

\subsection{Protocol Performance}\label{sec:results}
\begin{figure}[t]
\centering
\includegraphics[width=0.8\columnwidth]{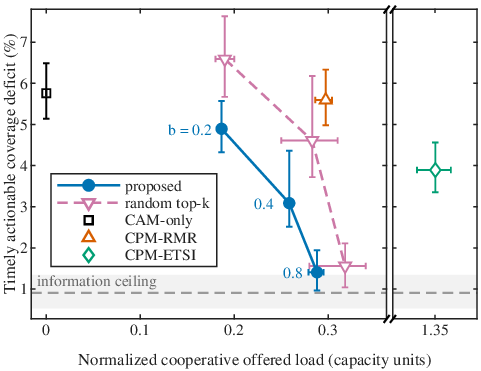}
\caption{Timely actionable coverage deficit versus normalized cooperative
offered load at density $1.5\,k_c$ and 80\,\% penetration. Shaded band:
oracle information ceiling. Error bars: arm-wise 95\% confidence
intervals, not pairwise comparisons.}
\label{fig:pareto}
\end{figure}

Fig.~\ref{fig:pareto} shows the deficit versus load frontier at density
$1.5\,k_c$ and 80\,\% penetration. CAM-only leaves a deficit of 5.8\%.
CPM-RMR spends 0.30 capacity units yet reaches only 5.6\%, while
unbudgeted CPM-ETSI needs 1.35 units for 3.9\%. The proposed selection
lowers the deficit monotonically with the allowance, from 4.9\% at
$b{=}0.2$ to 1.4\% at $b{=}0.8$. Random admission trails at every
allowance and is Pareto-dominated at the two upper ones. At $b{=}0.2$
random admission (6.6\%) is worse than CAM-only, since its airtime adds
little coverage while degrading beacon delivery. At $b{=}0.8$, whose offered load is comparable to CPM-RMR, the hosted twin cuts the deficit fourfold and comes within half a percentage
point of the oracle ceiling, while $b{=}0.4$ already reaches 3.1\% at a
load below CPM-RMR. The binding constraint migrates with the allowance. The recorded dual price averages $\lambda_{\mathcal{R}}\!\approx\!0.007$ at $b{=}0.2$, where the per-host allowance binds while the corridor-mean load, digest included, stays at 0.19, below $\Gamma_{\mathcal{R}}$, but $\lambda_{\mathcal{R}}\!\approx\!8$ at $b{=}0.8$, where the congestion price takes over and self-limits the load to 0.29, just below $\Gamma_{\mathcal{R}}$, although the allowance alone would not bind. At nearly equal occupancy, the proposed scheme spends 0.22 capacity units on reports net of the digest
against 0.30 for CPM-RMR, 27\% less reporting airtime for a fourfold
lower deficit.

\begin{figure}[t]
\centering
\includegraphics[width=0.9\columnwidth]{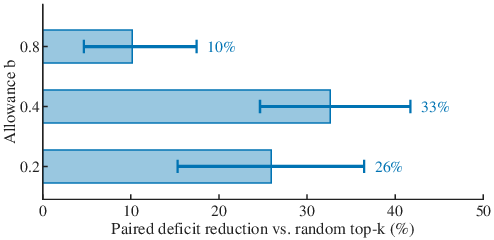}
\caption{Paired per-seed deficit benefit of the proposed selection over
random admission at the same per-host allowance. At the matched-load
allowance $b{=}0.2$ the gain isolates the selection criterion. The upper
allowances compare admission policies. Error bars are paired per-seed
95\% confidence intervals.}
\label{fig:selectivity}
\end{figure}

Fig.~\ref{fig:selectivity} quantifies the criterion's contribution. The
paired per-seed deficit reduction relative to random admission is 26\% at
$b{=}0.2$, 33\% at $b{=}0.4$, and 10\% at $b{=}0.8$, all intervals
excluding zero (the ordering between $b{=}0.2$ and $b{=}0.4$ is indicative only). At $b{=}0.2$ the two arms offer identical load, which gives the most direct estimate of the criterion alone. At $b{=}0.8$ random admission, which ignores price, overshoots
$\Gamma_{\mathcal{R}}$, so the 10\% mixes criterion and price effects.
So conditional scoring determines how effectively a given allowance is spent.

\begin{figure}[t]
\centering
\includegraphics[width=.9\columnwidth]{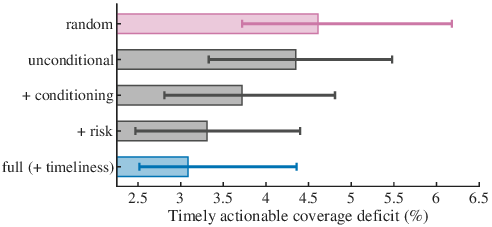}
\caption{Factor ablation of the admission score \eqref{eq:score} at
per-host allowance $b=0.4$, where each row adds one factor. The deficit
axis does not start at zero. Error bars are arm-wise 95\% confidence
intervals.
Adjacent steps are assessed by the paired per-seed differences in the
text.}
\label{fig:ladder}
\end{figure}

Fig.~\ref{fig:ladder} decomposes the criterion at $b{=}0.4$. Ranking by
absolute coverage improves little over random admission (4.4\% vs.\
4.6\% in arm means), since wide views overlap heavily at high
penetration. Conditioning on what the twin already holds is the largest single step in the mean, a paired per-seed reduction of 0.65\,pp $[0.32, 0.98]$. Risk weighting adds 0.39\,pp $[0.13, 0.68]$, and the timeliness factor adds 0.21\,pp $[-0.02, 0.41]$, an interval that includes zero, consistent with the $T_{\mathrm{cyc}}$ coupling of Sect.~\ref{sec:model} being nonbinding at this allowance. The offered load falls in step from 0.29 to 0.26, so each refinement achieves more coverage with less airtime, and the conditioning term carries the largest share of the criterion gain.

\balance
\section{Conclusion}\label{sec:conclusion}
This letter posed contributor admission for a mobile regional DT as a set-level conditional-coverage decision that per-sender mitigation cannot make, because only the fusion host knows what the region lacks. A beacon-borne coverage digest exposes candidate coverage before any detailed transmission, and a congestion-priced protocol admits non-redundant contributors within the cycle deadline. In the evaluated
80\%-penetration corridor the hosted twin cut the timely actionable
coverage deficit fourfold versus CPM-RMR at comparable load, and at the matched tight-budget point the criterion alone improved the deficit by
26\% over random admission. Broader density and penetration regimes,
adaptive hysteresis thresholds, host placement under non-uniform traffic, and lower-rate digests remain future work.


\end{document}